\documentclass{article}

\usepackage{PRIMEarxiv}
\usepackage{geometry}
\usepackage{booktabs}
\usepackage{etoolbox}

\providecommand{\keywords}[1]{\par\vskip 0.5em\noindent\textbf{Keywords:} #1}
\AtBeginEnvironment{tabular}{\small}

\usepackage[utf8]{inputenc} 
\usepackage[T1]{fontenc}    
\usepackage{hyperref}       
\usepackage{url}            
\usepackage{booktabs}       
\usepackage{amsfonts}       
\usepackage{nicefrac}       
\usepackage{microtype}      
\usepackage{fancyhdr}       
\usepackage{graphicx}       
\graphicspath{{media/}}     
\usepackage{mathtools}
\usepackage{amsthm}
\usepackage{enumitem}
\usepackage{amssymb}

\usepackage[font=small,labelfont=bf]{caption}
\usepackage{tikz}
\usetikzlibrary{positioning,shapes.misc,arrows.meta}
\usepackage{lineno}

\title{Federated Computing as Code (FCaC)\\
Sovereignty-aware Systems by Design}

\author{
  Enzo Fenoglio\thanks{Corresponding author.}, ~Philip Treleaven\\
  Dept. Computer Science \\
  University College London\\
  66-72 Gower Street, London WC1E 5EA, UK\\
  \texttt{\{e.fenoglio, p.treleaven\}@ucl.ac.uk}
}

\begin{document}
\maketitle

\begin{abstract}
Federated computing (FC) enables collaborative computation—such as machine learning, analytics, or data processing—across distributed organizations while keeping raw data local. Built on four architectural pillars: (a) distributed data assets (local control), (b) federated services (in-place computation), (c) standardized APIs (interoperability), and (d) decentralized services (multi-party control), FC is designed to support sovereignty-preserving collaboration. However, federated systems spanning organizational and jurisdictional boundaries lack a portable mechanism for enforcing sovereignty-critical constraints. They depend on runtime policy evaluation, shared trust infrastructure, or institutional agreements that introduce coordination overhead and offer limited cryptographic assurance.\textit{ Federated Computing as Code} (FCaC) is a declarative architecture that addresses this gap by compiling authority and delegation into cryptographically verifiable artifacts rather than relying on online policy interpretation. Boundary admission becomes a local verification step rather than a policy decision service. FCaC separates what we term constitutional governance (permission to execute and delegate across sovereignty boundaries) from procedural governance (context-dependent procedures during execution). Admission is validated locally and statelessly with respect to online policy services at execution boundaries using proof-carrying capabilities, while stateful services may still implement post-admission controls such as Attribute-Based Access Control (ABAC), risk scoring, quotas, and workflow state. FCaC introduces Virtual Federated Platforms (VFPs), which combine Core, Business, and Governance contracts through a cryptographic trust chain: Key Your Organization (KYO), Envelope Capability Tokens (ECTs), and proof of possession (PoP). We show the approach in a proof-of-concept setting on a cross-silo federated learning workflow using MNIST as a surrogate workload to validate the admission mechanisms and release an open-source proof of concept showing envelope issuance, cryptographic boundary verification, and envelope-triggered training.\\

Code and artifacts: \url{https://github.com/onzelf/FCaC-MNIST}.
\end{abstract}


 \section{Introduction}
Scientific research, healthcare innovation, and industrial optimization increasingly depend on collaborative analysis of data distributed across organizational and jurisdictional boundaries. Researchers need to train machine learning models on patient records held by multiple hospitals. Supply chains require analytics across suppliers without exposing proprietary information. Financial services face similar pressures since institutions must detect fraud patterns and perform risk and compliance analytics while preserving customer privacy and satisfying strict regulatory and audit obligations. These scenarios share a common requirement. They need to enable computation over distributed data while respecting sovereignty, privacy, and regulatory constraints. Three established architectures are commonly used to support multi-partner collaboration and data-intensive analytics:

\begin{itemize}[topsep=0.5pt, leftmargin=1.8em]
\item \textbf{Shared Database Platforms (Data Lakes):} Centralized repositories that aggregate data to enable broad exploration and integration. They are effective for internal analytics, but privacy regulations and sovereignty constraints often preclude their use for sensitive datasets, particularly in healthcare and cross-border scenarios~\cite{Hai_2023}.

\item \textbf{Trusted Research Environments (TREs):} Secure, controlled digital 
environments where authorized researchers can access and analyze sensitive datasets under strict governance. TREs are widely deployed in healthcare but typically operate within a single organizational boundary, limiting their ability to integrate datasets across multiple institutions~\cite{TRE_2022}.

\item \textbf{Federated Computing (FC):} A decentralized paradigm that enables collaborative computation across distributed organizations while keeping raw data local. FC provides orchestration infrastructure connecting data holders and analytics services (and, where applicable, controlled execution environments such as TREs), facilitating access to data and algorithms without centralization. FC supports multi-partner, multi-discipline, and multi-dataset analytics for preserving data sovereignty~\cite{Fenoglio_2025}.
\end{itemize}

Federated Computing (FC) has been discussed in the literature primarily as a collection of distributed computing mechanisms enabling collaboration across organizational boundaries, often in the context of federated learning~\cite{yang_2019} or decentralized system architectures. The survey by Schwermer et al.~\cite{schwermer_2024} provides a comprehensive taxonomy of technical components, extensions, and deployment models in this space. In contrast, the notion of Federated Computing adopted in this work originates from a different perspective. We introduced Federated Computing (FC) as a business infrastructure~\cite{Fenoglio_2025} and later refined it for sovereignty constraints~\cite{fc_rsos_2025}, where the original pillars had been recast as architectural principles. This view defines a framework for composing interoperable federated platforms aligned with organizational constraints, regulatory requirements, and collaboration agreements. The present paper builds on that perspective and introduces Federated Computing as Code (FCaC), which operationalizes the four principles by transforming sovereignty-critical constraints into locally verifiable conditions enforced at execution boundaries.

FCaC follows the later principle-based refinement of FC~\cite{fc_rsos_2025}, under which the original pillars are treated as architectural principles:
\begin{enumerate}[topsep=0pt, leftmargin=1.8em]
    \item \textbf{Distributed Data Assets:} Data remains distributed under the ownership of the originating entity (user, organization, or jurisdiction), to preserve privacy, enable regulatory compliance, and remove central points of failure. 

\item \textbf{Federated Services:} Computation moves to data and not the opposite. By bringing algorithms to distributed data sources, federated services enable collaborative analytics and model training while maintaining data locality.

\item \textbf{Standardized APIs:} Secure, standardized communication protocols between distributed components facilitate interoperability across heterogeneous environments, abstracting underlying infrastructure complexities. 

\item \textbf{Decentralized Governance:} Multi-party governance without a central authority enables verifiable interactions through implementation-agnostic mechanisms. It expresses the architectural requirement to distribute policy control and enforcement across independently governed participants, and not simply to decentralized infrastructure services.
\end{enumerate}
These principles address the technical challenges of distributed computation. Cross-domain deployments, however, introduce a different problem: no central authority can enforce delegation constraints across independent organizations. Participants retain autonomy, and trust cannot be assumed across boundaries. Traditional mechanisms—access control lists, policy engines, and identity providers—were designed for single administrative domains and depend on shared trust infrastructure or centralized coordination. Those assumptions break down when parties operate under different legal regimes and jurisdictional constraints.

This paper outlines the architectural scope of \textit{Federated Computing as Code (FCaC)}, a declarative modeling framework for expressing deployment, operation, and coordination of federated infrastructures through composable contracts. The \textit{as-code} paradigm automates processes that would otherwise be performed manually, improving consistency, repeatability, and version control. The notable examples include \textit{Infrastructure as Code (IaC)}~\cite{Terraform}, which codifies infrastructure resources, and Policy as Code~\cite{Chuprikov_2025}, which encodes security and compliance rules for automated enforcement. FCaC extends this logic to federated admission and delegation by encoding authority, rights, and delegations as verifiable artifacts instead of procedural agreements. Moreover, FCaC is built around a strict separation between admission (permission to execute and delegate) and post-admission operational control (context-dependent procedures during execution). It integrates deployment, execution, and boundary verification through \textit{Core, Business}, and  \textit{Governance} contracts (Section~\ref{ssec:cbg}), linking how systems are built, operated, and checked across organizational boundaries.

\subsection{Motivation: Governance Beyond Policy Engines}
Modern DevSecOps pipelines have achieved a high degree of automation in deployment, configuration, and monitoring. Infrastructure as Code, continuous delivery, and automated compliance checks have significantly reduced operational friction within organizational boundaries. Governance, however, remains largely procedural, mediated through access lists, policy engines, identity brokers, contracts, and audit processes that require continuous interpretation and coordination at runtime~\cite{hu_2014}. These mechanisms are effective within a single administrative domain, but they do not provide portable, auditable admission across independent authorities. Once data or computation crosses an organizational boundary, admission semantics depend on shared trust assumptions, synchronized policy state, or centralized intermediaries. As a result, enforcement becomes opaque, brittle, and difficult to verify after the fact.

Policy engines, RBAC matrices, and platform-specific onboarding workflows remain valuable operational tools, but they do not solve the cross-silo problem. In federated settings, authorization semantics must survive heterogeneity in stacks, vendors, and jurisdictions, and must remain explainable under external scrutiny. The issue is therefore not whether a local platform can evaluate policy correctly, but whether cross-boundary admission can remain portable, verifiable, and governance-preserving when no shared runtime authority can be assumed. The following failure modes motivate FCaC’s design:

\begin{itemize}
\item \textbf{Audit Reconstruction:} A regulator, auditor, or internal risk function asks: \emph{Why was this operation allowed, under which rule, at that moment, and who had stop authority?} If the answer is ``check the platform configuration and logs,'' then the decision must be reconstructed after the fact rather than verified directly. In practice, such reconstruction is brittle: policies evolve, logs may be incomplete or siloed, and the linkage between a request, the evaluated rule set, and the resulting boundary decision is often implicit~\cite{Cotrini_2018}. FCaC addresses this by turning execution into a proof-carrying act in which the requester presents signed artifacts encoding the admissible operation, scope, and validity window, and the verifier records a deterministic decision trace bound to those artifacts. The result is an auditable object---what was checked, against which signed constraints, under which trust anchors---rather than an inference from mutable platform state.

\item \textbf{Federation Drift:} Partners, vendors, and toolchains change. RBAC settings, policies, and onboarding workflows do not compose across stacks, so admission semantics are reinterpreted at every boundary~\cite{Selvanathan_2019}. The result is gradual drift toward bilateral exceptions, local translations, and operator-dependent trust. What begins as a federation degrades into a collection of ad hoc integrations. FCaC treats admission semantics as portable by construction: each participant establishes authority locally and issues signed capabilities whose constraints can be verified elsewhere without reinterpretation. Admission therefore becomes independent of any single vendor policy engine or centralized coordinator, because the enforcement surface consumes standardized, signed artifacts rather than platform-specific policy state.

\item \textbf{Derived Outputs:} Training is not the only governed act. Metrics, embeddings, predictions, and other derived outputs are also governed assets, each with its own distribution risk and regulatory profile. If participation in a consortium is treated as implicitly authorizing access to such outputs, sensitive or valuable information may leak even when raw data never moves~\cite{Shokri_2017,Fredrikson_2015}. FCaC makes these outputs explicit in the contract model by binding capability constraints to both operations and outputs (e.g., export, query, predict), together with time bounds and delegation limits. Access to derived artifacts must therefore be explicitly minted, presented, and verified, rather than inherited implicitly from participation in training.
\end{itemize}

These failure modes are structural rather than symptoms of immature tooling or incomplete standardization. Traditional control mechanisms often rely on runtime interpretation of context, intent, and mutable state to decide whether an action should be permitted. In cross-domain settings, that dependence makes admission fragile and difficult to audit. Organizations therefore resort to auxiliary mechanisms such as manual agreements, audits, legal contracts, and centralized coordinators to bridge the gap. These mechanisms scale poorly, offer limited cryptographic guarantees, and cannot reliably prevent unauthorized downstream use once data or computation leaves the originating domain. FCaC starts from the observation that some constraints are mechanically enforceable, while others remain inherently contextual. It isolates the subset that can be enforced deterministically and compiles authority, delegation, and capability scope into cryptographically verifiable artifacts that accompany each request. Cross-boundary enforcement is therefore carried by the request itself rather than reconstructed from mutable platform state. In this sense, FCaC shifts cross-domain governance from runtime policy interpretation to verification of signed constraints fixed at design time.

The rest of the paper is organized as follows. Section~2 reviews related work and positions FCaC relative to existing access-control and governance technologies. Section~3 describes FCaC’s architectural foundations. Section~4 presents FCaC’s central claim through the separation between constitutional and procedural governance. Section~5 introduces the cryptographic trust chain. Section~6 describes FCaC guarantees at the boundary and its threat model. Section~7 presents typical use cases. Section~8 discusses an illustrative example and the accompanying proof of concept. Section~9 discusses FCaC deployment models. Section~10 concludes.

\section{Related Work and FCaC Positioning}
FCaC relates to several established bodies of work in authorization, distributed governance, and security architecture. It does not compete with these approaches. It introduces an architectural boundary that constrains and composes them in federated, cross-organizational settings. This section situates FCaC within existing technologies and clarifies the assumptions under which they operate.

\paragraph{Identity and Policy-Centric Authorization Systems:}
A large class of authorization systems relies on identity and policy-centric decision making. Role-Based Access Control (RBAC)~\cite{sandhu_1996}) and Attribute-Based Access Control (ABAC)~\cite{Singh_2021,hu_2014} formalize authorization as policy evaluation over identities and attributes ~\cite{Jyosthna_2024}. Contemporary implementations often employ centralized, or logically centralized, authorization services that evaluate policies at runtime. Examples include OAuth 2.0 and OpenID Connect authorization servers ~\cite{rfc6749}, policy engines such as XACML ~\cite{oasis_2013}, or large-scale authorization systems like Google's Zanzibar ~\cite{zanzibar}, which provides global consistency by maintaining a shared authorization graph. More recently, Policy as Code approaches ~\cite{Chuprikov_2025} brought declarative specification to policy management. Systems such as Open Policy Agent (OPA)~\cite{opa_docs} and AWS Cedar~\cite{cedar_2024} enable policies to be defined as versioned code, evaluated consistently across infrastructure, and integrated into CI/CD pipelines. These systems advance auditability, testing, and reproducibility within organizational boundaries. FCaC shares the declarative intent and targets cross-organizational scenarios in which authorization cannot depend on a shared control plane. Identity and access management systems (IAM)~\cite{Sharma_2015} address authorization as a runtime decision problem. Requests are evaluated against mutable policy state and contextual attributes, typically through a logically centralized decision service or a shared authorization substrate. FCaC differs in targeting cross-domain settings where shared policy state or centralized decision services cannot be assumed, shifting boundary admission from online evaluation to local cryptographic verification.

\paragraph{Zero Trust Architecture:} 
Zero Trust Architecture (ZTA)~\cite{ZTA} is a security architecture that emphasizes continuous verification and explicit access decisions based on identity, device posture, and contextual signals. In practice, ZTA combines strong enforcement at the access point with a decision and telemetry ecosystem that depends on continuously updated context. FCaC is compatible with ZTA and addresses a different failure mode. ZTA strengthens enterprise enforcement, whereas FCaC makes cross-domain authority and delegation portable through verifiable artifacts checked locally at the boundary. ZTA mechanisms remain valuable inside each domain for operational controls, monitoring, and response.

\paragraph{Capability-Based and Proof-Carrying Authorization:}
Capability-based security has a long history, from early operating systems to language-based security models ~\cite{dennis_1966,miller_2003}. Subsequent work explored proof-carrying authorization and delegation using signed statements and tokens ~\cite{appel_1999, lampson_1992}. Modern token-based mechanisms, including JSON Web Tokens (JWT) ~\cite{rfc7519} and proof-of-possession extensions such as DPoP ~\cite{rfc9449}, provide practical realizations of these ideas. FCaC adopts this lineage and specializes it for sovereignty constraints by treating capabilities as derived artifacts from declarative governance contracts. It uses proof-carrying requests to support boundary decisions, while reserving context-dependent enforcement for post-admission control within each domain.

\paragraph{Governance Frameworks and Procedural Approaches:}
Distributed systems often rely on contractual agreements, audits, compliance workflows, and organizational trust arrangements to manage obligations that cannot be enforced mechanically. In data governance and federated learning, many requirements are handled through institutional arrangements, policy-compliance checks, and \textit{post-hoc} accountability mechanisms~\cite{rieke_2020, kairouz_2021}. These approaches are appropriate when obligations depend on context, interpretation, or organizational judgment. FCaC separates boundary verification from the contextual controls that remain inside backend systems and organizational processes. Existing frameworks therefore remain applicable under FCaC, but they apply after the request has crossed the boundary.

\paragraph{Federated and Decentralized Authorization Systems:}
Several systems address authorization across organizational boundaries through federation or decentralization. SAML ~\cite{cantor_2005} and federated identity protocols enable single sign-on across domains but require participants to trust designated identity providers. Decentralized identifier (DID) ~\cite{Sporny_2022} and verifiable credentials provide cryptographically verifiable identity and claim mechanisms without centralized authorities. Systems such as UCAN~\cite{UCAN_2022} extend this space toward decentralized authorization through delegable capability chains. FCaC shares the reliance on cryptographic proofs and local verification, but targets the broader problem of governed admission of cross-domain computation at execution boundaries rather than authorization alone.  

\paragraph{Blockchain and Distributed Ledger Systems:}
Blockchain-based systems and smart contracts ~\cite{buterin_2014} provide decentralized coordination through consensus and cryptographically verified state transitions. These systems reduce single points of failure and enable coordination among mutually distrustful parties through shared ledger state.  FCaC does not require participants to maintain consensus over a shared global state. Boundary decisions are checked locally from presented artifacts. Blockchain-based systems can still be used within FCaC Core and Business contracts as mechanisms for state management, coordination, and procedural governance~\cite{Soltani_2022}. For example, a federated platform can use smart contracts for resource allocation, payment settlement, or audit trails. FCaC’s constitutional layer constrains who may invoke such mechanisms and under which delegations. Architecturally, FCaC uses local proof validation for boundary admission across federated domains, whereas blockchain provides consensus-based state management within execution domains.

\paragraph{Dataspace Protocols and European Standardization:} Emerging dataspace initiatives provide an important adjacent point of comparison. In particular, the Eclipse Dataspace Protocol (DSP) defines schemas and protocols for interoperable data sharing, agreement negotiation, and data access across participants in a dataspace~\cite{DSP_2025}. FCaC focuses on cryptographically verifiable boundary control, where DSP focuses on exchange workflows and usage-governed interoperability. In fact, authority, delegation, and possession are compiled into locally checkable artifacts that determine whether execution may proceed. This distinction may become increasingly relevant as European standardization around data spaces matures, including the work of CEN/CENELEC JTC 25 on data management, data spaces, cloud, and edge~\cite{JTC25_DataAct_2025}.

\subsection{Positioning FCaC Relative to Existing Access-Control and Governance Technologies}
\begin{figure}[ht]
    \centering
    \includegraphics[width=1.0\linewidth]{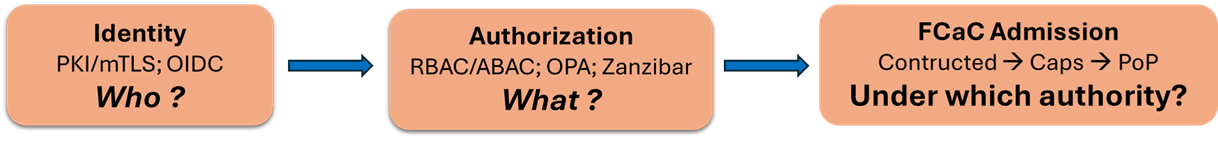}
\caption{FCaC Admission: identity establishes \emph{who}, authorization defines \emph{what}, and FCaC binds \emph{under which authority} via locally verifiable artifacts.}
\label{fig:fcac-positioning}
\end{figure}

To clarify where FCaC sits relative to existing approaches, Figure~\ref{fig:fcac-positioning} organizes the landscape around three questions that are often conflated in practice: \emph{(1) Who are you?} \emph{(2) What are you allowed to do?} and \emph{(3) Under which authority are you allowed to do it?} The first concerns identity and claims; the second concerns permissions under an adopted policy or relationship model; the third concerns the provenance and cross-domain validity of an operation. 

\begin{enumerate}[topsep=1.5pt, leftmargin=2.em]
    \item Identity technologies such as passwords, PKI/mTLS, SAML/OIDC, and verifiable-credential schemes establish control of an identifier and, in some cases, presentable attributes. They answer who the requester is, but do not by themselves determine which computations or data uses are admissible under a sovereignty-aware governance model. Their trust anchors remain institutional or registry-based, and the meaning of claims is external to the credential format.

    \item Mainstream access-control mechanisms---RBAC/ABAC, IAM policies, OAuth scopes, and policy agents such as OPA---address the second question by enumerating permitted actions under a current policy. Relationship-based authorization systems inspired by Zanzibar (e.g., SpiceDB~\cite{spicedb}, Keto~\cite{keto}) generalize this approach via authorization graphs. These systems are effective within a managed ecosystem, but their semantics are typically platform-bound since authorization depends on shared policy or relationship state and trusted runtime infrastructure, and audit often reduces to log inspection and reconstruction while we need a portable, cryptographically verifiable boundary record.

    \item FCaC addresses the third question: \emph{under which authority} a given operation is permitted across domain boundaries. It does so by expressing authority, delegation, and capability scope through portable cryptographic artifacts that can be verified locally at the boundary.
\end{enumerate}

\section{Architectural Foundation of FCaC}
\label{sec:archi}
\subsection{The Four Architectural Principles}
FCaC builds on the four architectural principles of Federated Computing as formulated in its sovereignty-oriented refinement~\cite{fc_rsos_2025}. These principles—distributed data assets, federated services, standardized APIs, and decentralized services—define the conditions under which distributed computation can remain sovereignty-preserving across organizational boundaries.

\begin{table}[ht]
    \centering
    \begin{tabular}{|l|p{5.5cm}|p{5.7cm}|}
    \hline
     \textbf{ Principles}   & \textbf{Implication} & \textbf{FCaC realization} \\ \hline
     \hline
\parbox[p]{11em}{Distributed data assets\\ (local control)}
  & Data remains in-domain; only bounded computation is permitted to reach it  
  & Boundary verification enforces capability scope so only permitted computations can execute against in-domain assets \\
\hline
\parbox[p]{11em}{Federated services\\ (compute-to-data)}
  & execution occurs inside domains; requests must be safely admitted without central coordination
  & The VFP verifier checks requests through locally verifiable proof-carrying artifacts; execution then proceeds within Core. \\
\hline
\parbox[p]{11em}{Standardized APIs\\ (interoperability)}
  & Cross-domain requests must be interpretable uniformly across heterogeneous stacks 
  & Governance contracts compile to verifiable artifacts (ECTs with proof of possession realized through DPoP) whose semantics remain stable across local policy environments. \\ \hline
\parbox[p]{11em}{Decentralized governance\\ (multi-party enforcement)}
  & No single party is trusted to enforce federation-wide constraints; delegation must be verifiable 
  & The trust chain KYO → ECT → PoP provides portable delegation evidence; in the present implementation, PoP is realized through DPoP, and the same inputs yield the same decision at each site. \\
\hline
    \end{tabular}
    \caption{From FC principles to FCaC mechanisms. Each principle becomes an implementation constraint under sovereignty. FCaC realizes these constraints through boundary verification, proof-carrying artifacts, and the VFP trust chain. VFP is Virtual Federated Platform; KYO is Key Your Organization; ECT is Envelope Capability Token; DPoP is Demonstrating Proof-of-Possession (Section~5)}
    \label{tab:4principles}
\end{table}

The pillars were introduced to describe a meta-framework for executive decision-making~\cite{Fenoglio_2025}, whereas the principles recast that framework as implementation-oriented requirements for practitioners~\cite{fc_rsos_2025}. Taken together, these principles define the conditions under which distributed computation can remain sovereignty-preserving across organizational boundaries. FCaC builds on this principle-based formulation by showing how the orchestration layer required by federation can be specified and implemented in line with Infrastructure as Code (IaC) principles. In this way, FCaC turns architectural principles into deployable mechanisms. It specifies how control and enforcement are distributed across independently governed participants, not merely how infrastructure is distributed. Cross-boundary constraints are then realized as verifiable conditions enforced at execution boundaries.

\subsection{Virtual Federated Platforms (VFPs)}
Federated Computing as Code organizes federated systems through Virtual Federated Platforms (VFPs). A VFP defines the federated boundary within which infrastructure, services, business relationships, and cross-boundary constraints are specified declaratively while remaining operationally independent. It provides a common structure for federation without imposing centralized control or shared runtime state.

A VFP spans multiple administrative domains and execution environments. Its role is to define where cross-boundary interactions occur and where requests must be checked. It is not a centralized runtime or a shared service. Rather, it links independently operated systems through common contract conditions and locally verifiable control at the points where domains interact.

\begin{figure}[ht]
    \centering
    \includegraphics[width=0.5\linewidth]{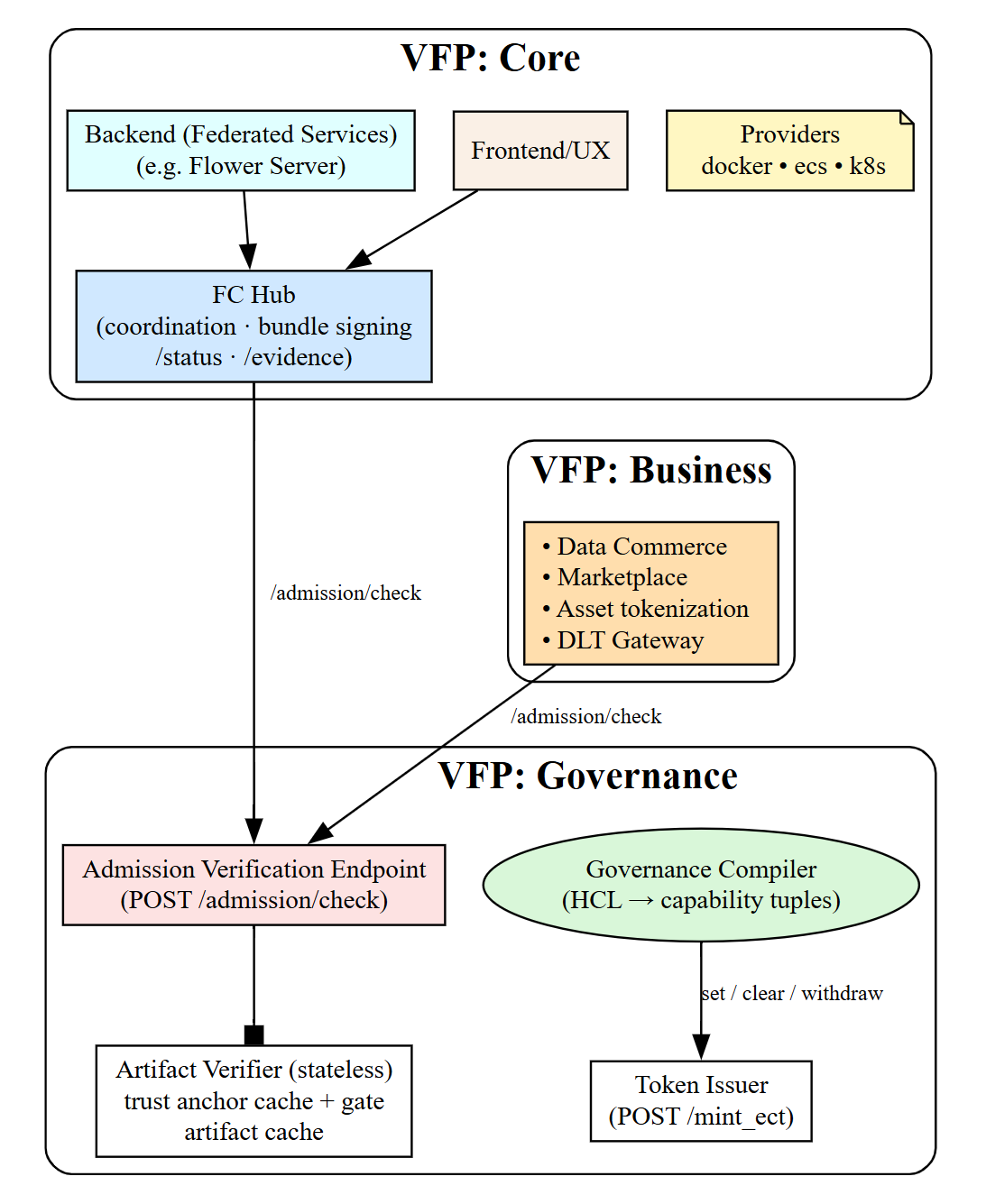}
    \caption{FCaC architecture across Core, Governance, and Business contracts.
FCaC binds Core services and Business interactions through proof-carrying artifacts derived from declarative contracts. At runtime, the boundary check is a verification step (artifact validity, possession binding, and capability match), not an online policy evaluation service. This yields a shared declarative model for infrastructure and policy automation, while keeping workflow and domain logic inside Core and Business components.}
    \label{fig:archi}
\end{figure}

The FCaC system architecture (Figure~\ref{fig:archi}) specifies each VFP through three declarative contract classes: \textit{Core}, \textit{Business}, and \textit{Governance}. Respectively, these capture local execution, cross-party interaction, and the conditions under which cross-boundary operations may occur.

\textit{Federated Services} continue to run inside their respective Core environments under local operational control. Core and Business components invoke verification against presented artifacts to determine whether a requested operation may proceed. Once that check succeeds, execution continues locally, including any application-specific controls, compliance checks, or workflow logic required by the participating system.

\subsubsection{Core, Business, and Governance Contracts}
\label{ssec:cbg}
The three contract classes introduced above specify complementary aspects of a VFP. Together, they define how federated infrastructures are instantiated, operated, and constrained across organizational boundaries.

\begin{itemize}[topsep=0.5pt, leftmargin=1.8em]
\item \textbf{Core Contracts} define the operational substrate of the VFP, specifying how services are deployed, configured, and executed within local infrastructure environments. They include infrastructure definitions, service topology, execution constraints, and operational integrations.

\item \textbf{Business Contracts} capture domain-specific intent and interaction semantics across parties. They encode terms of service, economic exchange, and application-level constraints relevant to cross-organizational workflows.

\item \textbf{Governance Contracts} specify allowed cross-boundary operations, authorities, and delegation structure. They compile into verifiable cryptographic artifacts used for local verification independent of deployment context. Trust anchors are distributed out of band and cached locally.
\end{itemize}

At a high level, a VFP can be read as extending a Terraform-style IaC module with business and cross-boundary control semantics. In this way, FCaC links infrastructure automation, service interaction, and cryptographic verification within a single federated structure (Section~\ref{sec:ConstProc}).
 
\section{Constitutional and Procedural Governance}  
\label{sec:ConstProc}
A central claim of FCaC is that not all governance constraints belong to the same enforcement layer. Some can be decided mechanically at the point where a request crosses an organizational boundary; others depend on local records, workflow state, temporal context, or institutional judgment and therefore remain inside the executing system. This section makes that distinction explicit. It introduces the separation between \emph{constitutional governance}, which determines whether cross-boundary execution may begin at all, and \emph{procedural governance}, which determines how an admitted operation is handled thereafter inside local environments. This distinction is the basis for the trust-chain mechanisms introduced in Section~5.

FCaC is built on a strict separation between two governance layers. The first is \emph{constitutional governance}, which defines the conditions under which an operation may enter the execution domain. The second is \emph{procedural governance}, which defines how an admitted operation is processed within local systems, workflows, and organizational processes. The terminology is borrowed from institutional law, but here it is given the following computational meaning:

\begin{itemize}[topsep=0pt, leftmargin=1.8em]
    \item \textbf{Constitutional governance} encodes rights and delegations as verifiable artifacts: \textit{who may be admitted, which capabilities exist, and under what authority and delegation chain they are valid}. It is enforced at the boundary and is designed to be stateless, deterministic, portable, and independent of local implementation details.
    \item \textbf{Procedural governance} covers the stateful and context-dependent controls that apply after admission, including workflow execution, rate limiting, scheduling, temporal conditions, compliance checks, and other domain-specific decisions. It remains local to the executing environment and depends on system state, operational context, and organizational process.
\end{itemize}
In operational terms, FCaC confines deterministic enforcement to the boundary. At the point where a request crosses into a new administrative domain, the verifier checks the presented artifacts for authority, delegation, and possession without relying on online policy evaluation or cross-party coordination. Once that check succeeds, execution proceeds inside Core services, where additional controls may still be applied according to local workflow state, compliance obligations, and application-specific logic. Therefore, the architectural point is not to replace procedural governance, but to separate it cleanly from boundary verification that remains portable and mechanically verifiable across independently governed domains.

About the terminology, \emph{admission} denotes the boundary decision to permit execution on the basis of proof-carrying artifacts, whereas \emph{authorization} refers to local access-control decisions inside the executing environment when describing related work in its own terminology.

\subsection{Constitutional Governance: Verifiable Commitments}
The constitutional layer covers the governance constraints that can be enforced deterministically at execution boundaries through cryptographically verifiable artifacts. It defines who may act, under which authority, and which operations may occur within a Virtual Federated Platform. Its role is to establish the boundary conditions under which execution may begin.

Two properties follow from this design. First, admission is deterministic, i.e., the same request, presented with the same artifacts and evaluated against the same trust anchors, yields the same outcome. Second, admission is portable, i.e., verification depends only on material carried with the request and locally available trust anchors, so enforcement can occur at any execution site without reliance on shared policy engines or centralized intermediaries.

The constitutional layer is intentionally narrow. It covers only those constraints that can be checked deterministically from the request and the presented artifacts. Questions that depend on workflow state, application context, or business semantics remain within Core services and are outside the boundary verifier's authority.

The result is a constitutive layer: it determines whether execution may proceed, but it does not coordinate workflows or prescribe how admitted operations must be carried out. Core and Business contracts govern what happens after admission within the bounds already defined and validated.

FCaC turns sovereignty-critical enforcement into a boundary property by grounding admission in verifiable commitments instead of retrospective inferences from logs or audits. Each admitted operation carries a cryptographic justification traceable to an issuing authority and an explicit delegation chain. This enables reproducible decisions across independently governed domains while preserving local autonomy after admission.

\subsection{Procedural Governance in Core}
Boundary enforcement in FCaC is limited to the decision to permit execution. Once a request has been admitted, execution proceeds inside Core services, where controls such as workflows, scheduling, rate limiting, monitoring, consent handling, lifecycle management, and compliance interpretation remain under local authority.

These controls are not secondary to the architecture; they are essential to the correct operation of the participating system. They govern how an admitted request is actually handled in practice, which internal workflow is triggered, which local records are consulted, which compliance obligations must still be checked, and which operational conditions may still cause the request to be rejected or deferred. Their evaluation may depend on mutable state, temporal conditions, external signals, local policy, or application semantics, and therefore cannot, in general, be reduced to deterministic verification based solely on proof-carrying artifacts.

For this reason, FCaC does not attempt to standardize or externalize these controls. Its role is narrower, limited to ensuring that sovereignty-critical constraints are enforced first at the boundary before execution enters the local environment. Procedural governance then determines how an admitted operation is processed inside the system. A Core service (e.g., Federated Learning)  may still reject the request on local grounds, apply additional restrictions, or enforce duties that arise only during execution.

The architectural point is therefore not to replace procedural governance, but to separate from boundary verification cleanly. The constitutional governance determines whether execution may begin across organizational boundaries; the procedural governance determines how execution unfolds thereafter within the participating domain. This separation preserves local autonomy, ensuring that cross-domain constraints are enforced through portable, cryptographically verifiable artifacts. It also determines what is compiled into capability artifacts and what remains inside backend logic as described in the next section.

\subsubsection{Policy tuples}
FCaC compiles only the subset of declarative clauses that can be enforced deterministically at the boundary into a finite set of \textit{capability tuples}, i.e., structured predicates over a small, fixed vocabulary (e.g., resource, action, and scope qualifiers) derived from the system policy. An example of the corresponding \textit{policy.json} artifact is given in Section~\ref{ssec:endpoints}, Figure~\ref{fig:policy}. Compilation is deterministic and performed at design time; it introduces no new semantics and does not consult runtime context. The tuple compiler extracts only the constitutional subset of the policy model; procedural rules, if present, remain with \textit{VFP:Core backend} logic and are not compiled into tuples. The resulting tuples are embedded in the ECT as signed claims that define what a request may do under which constraints. At verification time, the verifier checks artifact integrity and holder binding, then matches the request against the signed tuples.

\paragraph{Capability tuples.}
A \emph{capability tuple} is a structured predicate representing one admissible class of requests under the constitutional subset of the policy model. In schematic form, we write
\(
\tau = (r, a, s, c),
\)
where \(r\) denotes the protected resource, \(a\) the admissible action, \(s\) the relevant scope qualifiers, and \(c\) a finite set of signed caveats or restrictions (e.g., output class, audience, validity bounds, or delegation limits). Intuitively, a tuple states that a holder may request action \(a\) over resource \(r\) within scope \(s\), subject to the additional signed constraints collected in \(c\). ECT minting compiles declarative clauses into a finite set of such tuples, which are then embedded as signed claims and later matched against incoming requests during verification.
 
\subsection{Stateless Admission and Deterministic Boundary Decisions}
This subsection illustrates the distinction introduced above through two examples. In FCaC, the boundary verifier decides whether a request may proceed by checking the presented artifacts and locally cached trust anchors. Once that check succeeds, execution continues inside Core services, where local controls, workflow state, and domain-specific logic still apply. The contrast is therefore not between security and no security, but between what can be decided deterministically at the boundary and what must remain inside the executing system.

\textbf{Example 1 -- Cross-jurisdiction healthcare federation.} Consider a setting in which Hospital A permits a clinic to run a federated training job on dataset \texttt{D} under study identifier \texttt{S}. A request may proceed only if it presents a capability artifact that names the dataset and study scope, constrains the permitted operation to training, sets a validity window, and is linked through a verifiable delegation chain to a trusted issuer. These checks are decidable at the boundary. By contrast, controls such as consent validation for the relevant cohort, ABAC on internal attributes, rate limits, monitoring, and incident response remain inside Core services and are evaluated using local records and processes.

In a Virtual Federated Platform (VFP), this decision is made from verifiable artifacts expressing authority, delegation~\cite{rfc2693,rfc2692}, and proof of possession realized in the present implementation through DPoP~\cite{rfc9449}. The verifier checks signature chains, possession bindings, scope constraints, and freshness conditions using only the presented artifacts and local trust anchors. Given the same request and the same inputs, independent verifiers reach the same result. Missing or invalid proofs are rejected unconditionally, with no discretionary fallback path.

\textbf{Example 2 -- Temporal constraints.} The same distinction applies to time. Some temporal rules can be enforced directly from the current request, the current time, and the validity conditions carried by the presented artifacts. Examples include not-before times, expiry times, study windows, and approved hours of operation. An issuer may therefore mint a capability that permits training on dataset \texttt{D} for study \texttt{S} only between dates \texttt{T0} and \texttt{T1}, and only during approved hours in the issuer’s jurisdiction. Requests outside that window are rejected deterministically at the boundary.

Other temporal rules cannot be decided in this way because they depend on event history or sequencing. Obligations such as “\textit{after X, Y must occur within 24 hours}” depend on prior events, workflow progression, and local records. They therefore remain inside Core services, where monitoring, audit, and workflow state are available.

\section{The FCaC Cryptographic Trust Chain}
\label{sec:trustChain}
This section introduces the cryptographic trust chain that makes cross-domain boundary decisions portable across independently governed domains. The core idea is to represent sovereignty-critical governance as proof-carrying artifacts that can be validated mechanically at the execution boundary. The remainder of the section specifies how FCaC carries verifiable evidence of authority, delegation, and possession, and how these artifacts compose into a deterministic boundary decision.

The trust chain is composed of three elements: (i) \textbf{Key Your Organization (KYO)}, based on RFC 8705~\cite{rfc8705}, to anchor authority to a verifiable source; (ii) \textbf{Envelope Capability Tokens (ECTs)}, based on RFC 7519, RFC 7800, and RFC 7638~\cite{rfc7519, rfc7800, rfc7638}, to express admissible actions and delegation constraints; and (iii) \textbf{proof of possession (PoP)}, realized in the present implementation through DPoP based on RFC 9449~\cite{rfc9449}, to bind capability exercise to a concrete act of execution. Individually, these elements are insufficient. Taken together, they form a minimal basis for enforcing boundary verification at execution time (Figure~\ref{fig:trustchain}). For completeness in Appendix~\ref{app} we briefly recall the main RFCs employed in the FCaC trust chain.

 \begin{figure}[t]
    \centering
    \includegraphics[width=0.85\linewidth]{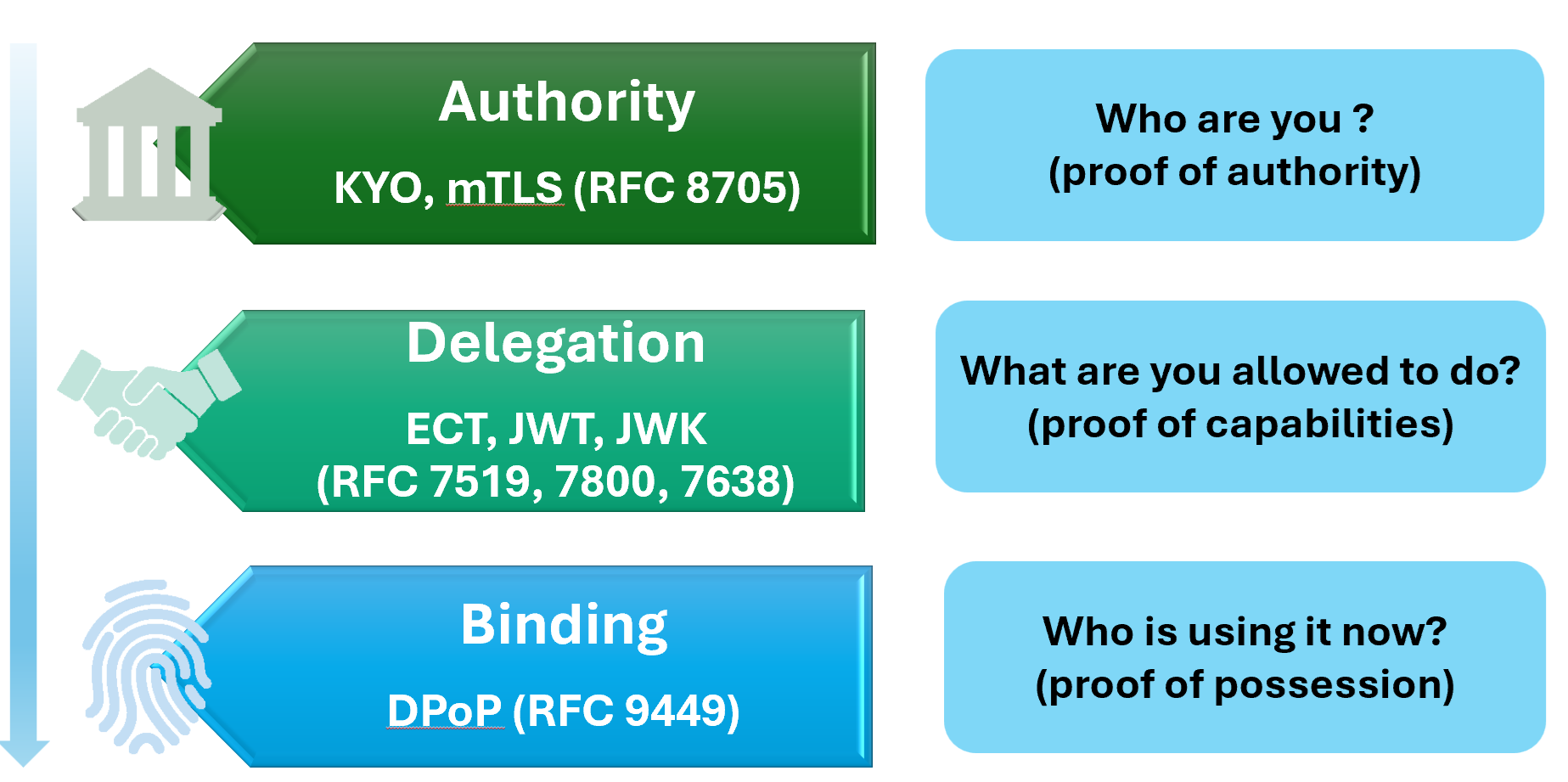}
    \caption{FCaC cryptographic trust chain: Key Your Organization (KYO), Envelope Capability Tokens (ECTs), and Proof-of-Possession (PoP)}
    \label{fig:trustchain}
\end{figure}

\subsection{Proof of Authority (KYO -- Key Your Organization)}
We use the acronym \textit{Key Your Organization} (KYO)\footnote{not to be confused with ``Know Your Owner/Ownership'' used in financial compliance} to bind the sovereign cryptographic identity of each participating organization within a Virtual Federated Platform. It answers the question \textit{"Who are you?"} at the organizational level. KYO is grounded in direct cryptographic proof of key ownership rather than in institutional identity providers or external trust authorities. In the intended deployment model, organizational credentials and authority-sensitive signing operations are anchored in a trusted execution environment (TEE), creating a protected enclave for key material and related operations~\cite{TEE_2022}. Organizations authenticate through mutual TLS (mTLS) handshakes~\cite{rfc8705} that demonstrate control over their private keys and associated digital credentials. This process establishes the organizational cryptographic root from which subsequent capability artifacts derive.

In KYO, legitimacy is not inferred from naming, registration, or institutional endorsement alone, but from the ability to produce verifiable cryptographic evidence. Authority is therefore treated as a property of key control since what matters is not only who claims authority, but who can prove control of the corresponding cryptographic material. Once validated, the organization's public key, together with minimal non-authoritative metadata, is recorded in the federation's trust registry. The registry serves only as a verification reference, allowing participants to check the provenance of issued capabilities against registered public keys without further coordination or runtime trust negotiation. Verifiers rely on locally cached trust anchors on the request path and therefore do not depend on the registry's availability at runtime.

KYO does not itself grant permissions or authorize execution. Its role is to establish the organizational cryptographic root from which issuer authority for ECT signing and holder-key verification for request execution are derived. Permissions, scopes, and delegation constraints are expressed through Envelope Capability Tokens (Section~\ref{ssec:ECT}), while request-time exercise is later bound through proof of possession (Section~\ref{ssec:pop}). This keeps the principal purely cryptographic across domain boundaries, i.e., requests are decided on the basis of signed artifacts and registered keys, not on UI-provided usernames or local application identifiers. KYO therefore separates the establishment of authority from its later delegation and exercise, providing an auditable organizational root of trust across jurisdictions without requiring a shared identity infrastructure.

\subsection{Proof of Capability (ECT -- Envelope Capability Token)}
\label{ssec:ECT}
The Envelope Capability Token (ECT) is the mechanism by which a participating organization expresses delegated capability in a portable, verifiable form. It answers the question \textit{"What are you allowed to do?"}. Whereas KYO establishes the organizational cryptographic root, the ECT carries the specific permissions, scopes, and delegation limits that may be exercised under that authority. The issuing authority examines an organization's declarative contracts, derives a minimal set of capability tuples, and encodes them as cryptographically signed artifacts. Each ECT therefore carries a scoped subset of the issuer's rules, expressing what may be done, under which constraints, and on whose delegated authority, without exposing the full specification or requiring it at runtime.

ECT minting is a deterministic \textit{tuple compiler} executed at design or issuance time. Declarative contracts are mapped to a finite set of capability tuples that capture admissible operations and constraints, including resource, action, scope, output restrictions (such as aggregation, PII, or contact), and non-functional caveats such as audience. These tuples are then embedded as signed claims in an Envelope Capability Token. Each ECT is bound to a specific holder and materialized as a JSON Web Token (JWT, RFC 7519~\cite{rfc7519}) carrying signed capability clauses and linked to a holder key via the JWT confirmation (\textit{cnf}) claim (RFC 7800~\cite{rfc7800}), using a JWK thumbprint (\textit{jkt}, RFC 7638~\cite{rfc7638}) to support proof of possession. The compilation step introduces no new semantics. It does not interpret intent, resolve context, or consult runtime state. Its role is to freeze capability constraints into a verifiable artifact suitable for local verification.

An ECT defines the capability boundary within which execution may proceed. Once minted, its scope and limits are fixed. They cannot be expanded by the holder, altered by backend services, or reinterpreted at runtime. In this sense, the ECT makes capability scope an invariant of execution rather than a policy decision deferred to the environment in which it is presented.

At runtime, the verifier is intentionally minimal. It validates the ECT signature and claims under the issuer's KYO key, checks the holder binding, and verifies that the resulting capability covers the requested operation. No policy parsing or external consultation is required; requests that fail these checks are rejected unconditionally.

ECTs therefore express scoped permission to attempt execution under explicitly defined constraints. They make cross-domain requests portable and locally verifiable without relying on shared policy engines or centralized trust infrastructure. Passing this check, however, does not guarantee successful execution. Once a request has entered Core services, it may still be rejected on the basis of workflow conditions, compliance obligations, local state, or other domain-specific controls. The ECT thus defines the outer limit of what may be attempted, while leaving the subsequent handling of the request to local systems. Because each ECT is traceable to an issuing authority and a specific contract version, it provides the basis for reproducible decisions across federated domains and prepares the ground for the final element of the trust chain: proof of possession, which binds capability exercise to a concrete act of execution.

\subsection{PoP -- Proof of Possession}
\label{ssec:pop}
The final element of the FCaC trust chain is proof of possession (PoP), which binds an issued capability to a concrete act of execution. It answers the question \textit{"Who is using it now?" }. KYO establishes the origin of authority. ECTs express admissible actions. PoP demonstrates that the entity presenting the capability for the current request controls the cryptographic key to which that capability is bound, so capability use cannot be detached from execution.

In the present implementation, this PoP layer is realized using DPoP~\cite{rfc9449}. For each request, the presenter constructs a DPoP proof by signing selected elements of the HTTP request with a private key under its control. This proof accompanies the request together with the relevant Envelope Capability Token. The verifier then validates the DPoP proof against the corresponding public key bound to the ECT, confirming that the request is being made by an entity in control of the key associated with that capability in that execution context.

DPoP establishes cryptographic possession, i.e., proof that the requester controls the key bound to the presented capability at the moment of use. By incorporating request-specific elements and freshness controls such as nonces, it provides replay resistance and helps prevent capability reuse outside the intended execution context. It therefore binds capability exercise to the concrete request without relying on bearer-token semantics alone.

This binding is essential to prevent capabilities from degenerating into transferable bearer tokens. An ECT alone expresses permission to attempt an operation, but without PoP, it could be replayed, proxied, or misused by unauthorized parties. PoP closes this gap by transforming capability use from a static claim into a live cryptographic act inseparable from execution.

Together, ECT and PoP enforce non-transferability by construction. Capabilities may still be delegated explicitly through the trust chain, but they cannot be exercised implicitly by intermediaries or copied across execution contexts without demonstrating control of the bound key. Each permitted operation is therefore accompanied by verifiable evidence that the requester controls the key bound to the capability at the time of execution.

\section{Cryptographic Accountability and Trust Containment}
This section describes what FCaC guarantees at the boundary, how those guarantees are maintained over time, and what remains outside its scope. We characterize the threat model for proof-carrying boundary verification, discuss trust continuity through key rotation and revocation, and summarize the failure modes that remain local or organizational by nature.

\subsection{Admission Control Semantics}
Admission control is implemented as a proof-carrying check over two artifacts: (i) an Envelope Capability Token (ECT) and (ii) a request-bound proof of possession, realized in the present implementation through DPoP. The ECT is a signed JWT (RFC 7519) carrying a finite set of capability tuples and bound to a specific holder key via the confirmation claim \textit{cnf.jkt} (RFC 7800, RFC 7638). DPoP binds that same holder key to a concrete execution event by signing request-specific parameters. A request is accepted only if the ECT verifies under the issuer's trust anchors, the proof of possession verifies against the holder binding, and at least one signed capability tuple covers the requested execution tuple. The result is a deterministic decision procedure whose inputs are explicit and whose outcome can be recorded as an auditable decision record. This composition defines the scope of FCaC guarantees at the boundary. It provides authenticity, possession, and capability coverage without requiring online policy evaluation or centralized enforcement. It does not resolve intent beyond what is encoded in the capability tuples, nor does it capture obligations that depend on off-chain process state. The rest of this section characterizes the threat model for this verification rule and the residual failure modes that remain local or organizational by nature. 

\subsection{Cryptographic Accountability Across Domains}
FCaC makes accountability more directly tied to execution rather than relying solely on retrospective inference from logs, audits, or institutional trust. Every operation that crosses a federated boundary carries verifiable evidence, so attribution is established at the moment of execution. This property follows from the FCaC trust chain. Key Your Organization (KYO) anchors capabilities to a sovereign authority. Envelope Capability Tokens (ECTs) express delegated capability and scope constraints. Proof of possession binds capability use to a concrete execution event. Together, these elements allow an accepted operation to be attributed cryptographically to an issuing authority, a specific capability, and the party exercising it at execution time. Because this evidence travels with the request and is verified locally using cached trust anchors, accountability remains portable across organizational and jurisdictional boundaries. Each execution environment can also verify under which contract version and delegation chain the request was accepted. Accountability does not imply that backend services are correct or benign. Core services may be buggy, misconfigured, or even adversarial, but within FCaC successful execution cannot exceed the authority established at the boundary. Operations lacking cryptographic justification are rejected regardless of backend intent. This differs from conventional approaches, in which attribution is inferred indirectly from access logs, identity assertions, or contractual arrangements, and often requires interpretation or institutional arbitration. By grounding accountability in verifiable artifacts, FCaC decouples assurance at the boundary from trust in internal operations. Participants need not trust each other's internal controls or operational discipline; they need only trust local verification of the presented artifacts, thereby clarifying cross-domain attribution and keeping delegated authority bounded.  

\subsection{Threat Model}
The threat model of FCaC focuses on unauthorized cross-boundary execution rather than on the correctness or availability of backend systems. FCaC does not attempt to prevent all classes of attacks in distributed systems; instead, it constrains a specific failure mode: the execution of operations that violate sovereignty or delegation constraints without valid cryptographic verification at the boundary. Under this model, an attack succeeds only if an operation that exceeds the declared capability constraints is executed without presenting valid cryptographic evidence at the boundary. By construction, FCaC aims to prevent the class of failures in which a sovereignty-violating operation may cross the boundary unless it has been cryptographically validated under the declared authority and delegation structure. This guarantee is independent of backend implementation choices, organizational trust assumptions, or deployment topology. It holds wherever local verification is performed against the same presented artifacts and trust anchors. Therefore, FCaC remains composable with conventional security controls and operational practices while enforcing a clear structural invariant at the federation boundary. It explicitly excludes several other threat classes from its scope. It does not address unauthorized disclosure or misuse of data after access has been granted, correctness of application logic, denial-of-service attacks, model integrity, or compliance obligations that require contextual or temporal interpretation. Such concerns remain the responsibility of backend systems and organizational processes. 

\subsection{Key Rotation, Revocation, and Trust Continuity}
\label{ssec:krot}
FCaC treats lifecycle control as part of trust continuity rather than as part of the core trust-chain construction. In the base design, cryptographic artifacts remain bounded by their declared validity windows and by the continued validity of the issuer key. A verifier therefore rejects expired artifacts deterministically, while key rotation prevents new capability artifacts from being minted under compromised or superseded keys.

More immediate withdrawal introduces a different class of mechanism. If a still-valid capability artifact must be invalidated before expiry, verifiers need additional revocation evidence distributed out of band and cached locally. The same applies to delegation links whose validity must be withdrawn before their natural expiration. These mechanisms improve responsiveness, but they also introduce operational state into the trust-distribution channel and therefore move beyond the minimal stateless design.

Three lifecycle cases are relevant in practice:
\begin{itemize}[topsep=0.5pt, leftmargin=1.8em]
    \item \textbf{Issuer key rotation or revocation:} withdraws the ability to mint new capability artifacts under a compromised or superseded key and changes which trust anchors are accepted during verification.
    \item \textbf{Capability revocation:} withdraws a specific artifact before expiry, requiring revocation evidence or, alternatively, sufficiently short-lived artifacts.
    \item \textbf{Delegation revocation:} withdraws a delegation link on which later artifacts depend, again requiring either revocation evidence or short-lived delegated artifacts.
\end{itemize}

The proof of concept implements bounded validity and key rotation. Immediate artifact revocation and delegation-specific revocation remain future work.

\section{Practical Implications and Use Cases}
FCaC provides a unifying framework for trustworthy collaboration across domains where coordination is typically difficult. It enables rights and obligations to travel with the computation rather than with the location of the data.

\subsection{Federated Research and Healthcare}
In federated healthcare research, hospitals must collaborate on shared analytics or machine-learning models without violating patient privacy or jurisdictional data laws~\cite{Elkourdi_2024}. Traditional approaches depend on access lists, VPNs, federated learning coordinators, and other intermediary control mechanisms~\cite{li_2020}, all of which require trust in shared operators or infrastructure. In this context, each hospital performs KYO to establish its authority, then authorized issuers mint ECTs that constrain permissible datasets, time ranges, and analytical purposes. Clinicians present proof of possession (e.g., DPoP) for the capabilities assigned to them, and each training or inference request is verified locally at the boundary. No raw data leaves institutional boundaries; only permitted updates (e.g., encrypted gradients) traverse the network.

\subsection{Industrial and Supply-Chain Data Exchange}
Modern manufacturing and logistics ecosystems require continuous data exchange among suppliers, integrators, and auditors while preserving commercial and regulatory boundaries. FCaC replaces brokered trust with cryptographically enforced sovereignty. Each participant establishes its authority through KYO, then compiles declarative governance rules into ECTs that specify which analytics or data views are permitted. An integrator, for instance, may receive an ECT allowing performance aggregation but not access to raw telemetry. Every analytic request includes proof of possession (DPoP), showing that the requester controls the key bound to that capability for the current request. The boundary verifier at each site verifies signatures locally. Admission succeeds only if the proofs align with the trusted keys and declared constraints. The result is a verifiable supply-chain fabric where data, computation, and policy coexist under the same cryptographic guarantees.

\subsection{Financial Services and Regulatory Analytics}
Financial institutions must collaborate on fraud detection, \textit{anti-money laundering (AML)} monitoring, and risk analytics across subsidiaries and counterparties to preserve customer confidentiality and meet strict regulatory and audit obligations. FCaC makes admission portable and locally verifiable across independently governed participants. Each institution establishes its authority through KYO, then mints ECTs that constrain admissible computations, data views, delegation scope, and validity windows. A bank, for example, may issue an ECT that allows the computation of aggregated risk indicators or alerts for a defined segment without exposing raw transaction records. Each request includes a DPoP proof-binding capability exercise for the concrete request. Boundary verifiers validate the presented artifacts locally. Admission succeeds only if signatures, bindings, and declared constraints hold. This yields cross-institution analytics with reproducible decisions and clear accountability, without shared policy engines or centralized intermediaries.

\paragraph{High-Frequency Trading (HFT):} A more latency-critical, yet equally regulation-driven, variant arises in electronic trading and \textit{High-Frequency Trading (HFT)}~\cite{Breckenfelder_2019}. In this setting, the key question is not only whether a strategy may run, but which version may execute, where it may be deployed, under which hard risk limits, and with what audit evidence. FCaC separates the ultra-low-latency execution path from the boundary-verification path. A strategy can be packaged as a sealed envelope whose identity is fixed by a hash over code, configuration, and optional model weights. An ECT then binds that identity to an approved execution scope, enforceable trading limits, and the evidence streams required for later review. At each execution site, the verifier checks the sovereignty envelope and the ECT for user requests. If the request is allowed, it releases short-lived runtime credentials. This mechanism  yields a portable, cryptographically auditable boundary layer for multi-party trading arrangements involving buy-side strategies, broker market access, and venue connectivity. The resulting decision record --- \textit{what ran, where, under which limits, and with what evidence} --- remains reproducible across parties and defensible in post-incident review without requiring a shared policy engine.

\section{An Illustrative Case Study}
This section grounds FCaC in an \emph{illustrative} cross-silo deployment that exercises the full trust-chain pipeline end to end. For reproducibility and ease of experimentation, we use MNIST as a surrogate workload. Concretely, we treat MNIST images as stand-ins for regulated clinical imaging assets (e.g., PET-CT slices) and use digit classes to emulate governance-relevant cohorts. Cohort labels (\texttt{EVEN\_ONLY}, \texttt{ODD\_ONLY}, \texttt{ODD\_PLUS}) define surrogate scope partitions in the PoC. They stand in for patient strata, study arms, or jurisdiction-specific cohorts in a simplified MNIST-based setting. These labels are encoded in the policy artifact through \texttt{scope.cohort}, propagated into the ECT at minting time, and checked again by the verifier when a request is presented. This setup allows us to demonstrate three properties: (i) cryptographically enforced request validation through ECTs and proof of possession, (ii) rejection of scope escalation when a request is tampered with to target a broader or different cohort than the one authorized, and (iii) separation between request verification and additional controls enforced later inside the federated service.

We deliberately keep the ML task simple while preserving the same trust-chain surface that would appear in a production clinical FL deployment: capability issuance, scope binding, boundary verification, and guarded service invocation. In the prototype, a training request is admitted only if it presents (i) a signed capability artifact that constrains scope, actions, and delegation limits, and (ii) proof of possession bound to the request. Requests that omit these artifacts, present invalid signatures, or exceed declared bounds are rejected deterministically at the boundary. Raw data never leaves its originating organization; only permitted outputs of admitted execution, such as model updates, are exchanged. Detailed implementation and deployment material is available in the FCaC repository on GitHub\footnote{~\url{https://github.com/onzelf/FCaC-MNIST}}, including contract examples, the deterministic tuple-to-token compilation pipeline, and deployment scripts for the complete end-to-end demonstration. 

We provide a description of the PoC interface and observed outcomes. We start from the two critical artifacts that are tested: (i) the \textit{sovereignty envelope} that is created via multi-party quorum (Bind → approvals → Active envelope), and (ii) the \textit{capability admission decisions} that is created with the Envelope Capability Token (ECT) bound to a DPoP proof (holder key binding) and evaluated against a \textit{policy.json} artifact. The PoC objective is to demonstrate verifiable compilation targets (policy artifacts, envelopes, attestable decisions) rather than optimize performance or implement a full production grade substrate.

\subsection{PoC endpoints and trust boundaries}
\label{ssec:endpoints}
This section describes the external endpoints exposed by the PoC and the points where trust is enforced. We identify where TLS is terminated, which identities are permitted on each route (hub versus organization administrators), and which attributes are treated as authoritative by the application (e.g., the forwarded mTLS subject DN). We then enumerate the minimal endpoint surface required to realize the two FCaC primitives used in this case study: (i) sovereignty envelope formation under quorum approvals, and (ii) proof-carrying requests verified locally. The goal is not to catalogue implementation details, but to make the enforcement perimeter explicit and reproducible for independent re-execution.

\paragraph{mTLS edge (nginx) and identity forwarding.}
The PoC places nginx at the trust boundary and terminates TLS there. Mutual TLS (mTLS) is used to authenticate calling parties before any request reaches the application. The verifier endpoint is authenticated with a PKIX X.509 v3 certificate whose service identity is carried in the \texttt{subjectAltName} (SAN) extension (e.g., \texttt{dNSName=verifier.local}), so hostname verification follows current PKI practice and not legacy CN-only matching~\cite{rfc9525,rfc5280}. Once the client certificate has been validated, nginx applies route-level identity constraints (for example, distinguishing Hub from Hospital-admin certificates) and forwards only the already verified identity attributes to the application. In this way, the application relies on identity facts established at the mTLS edge.

\paragraph{Envelope formation endpoints.}
The envelope creation flow uses the following endpoints behind nginx:
\begin{itemize}
    \item POST /beta/bind/init (hub-only)
    \item GET /verify-start (admin-only: Hospital A + Hospital B)
    \item GET /session/claim?code= (hub-only)
    \item POST /beta/bind/approve (hub-only; second approval reaches quorum)
\end{itemize}

\begin{figure}[ht]
    \centering
    \includegraphics[width=1.0\linewidth]{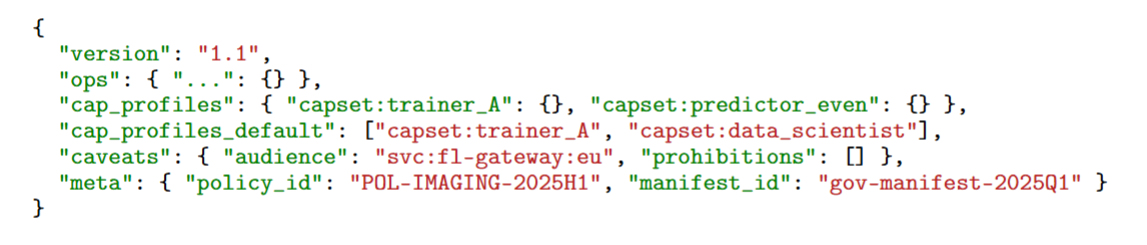}
    \caption{Minimal illustrative fragment of a policy.json file. The example shows the main policy elements used by FCaC—operation set, capability profiles, default capability assignment, caveats, and policy metadata—while omitting the additional fields and detail present in the full policy used in the proof of concept.}
    \label{fig:policy}
\end{figure}

\paragraph{Operational policy artifact: }
The PoC exposes admission constraints into a versioned policy artifact (\textit{policy.json}, v1.1) that provides the shared reference vocabulary for both capability issuance and boundary verification (Figure~\ref{fig:policy}). In practice, \textit{policy.json} defines the admissible operations, profiles, caveats, and other governance-relevant fields from which ECT claims are derived at minting time. The same artifact is then reused by the verifier at runtime to project an incoming request to a canonical tuple and to check whether that tuple falls within the signed bounds carried by the presented ECT and proof of possession. This arrangement keeps the semantics stable across issuance and verification, because issuer and verifier rely on the same declarative vocabulary. The artifact enumerates admissible operations (ops) (e.g., \textit{train\_petct\_A, predict\_petct\_even\_only}, \textit{export\_model\_params}) and defines named capability profiles (\textit{cap\_profiles}) that constrain which operations and scopes may be requested under a token (e.g., capset:trainer\_A, \textit{capset:predictor\_even}, \textit{capset:egress\_safe}). A default profile set (\textit{cap\_profiles\_default}) specifies the baseline granted profiles for a role when minting is not parameterized. The caveats block binds tokens to an intended service boundary (e.g., \textit{"audience": "svc:fl-gateway:eu"}) and may include explicit prohibitions. Finally, meta provides traceability (\textit{policy\_id}, \textit{manifest\_id}) and documentary notes on constraints handled outside the ECT (e.g., deployment routing or upstream quality gates). In \textbf{Test\#2} and \textbf{Test\#3}, the verifier uses \texttt{policy.json} as the reference vocabulary (ops, profiles, caveats) to project each incoming request to a canonical tuple, and then decides \texttt{ALLOW/DENY} by checking that tuple against the signed bounds carried in (ECT + DPoP).

\subsection{Evidence: acceptance tests}
We present the PoC through three acceptance tests: \textbf{Test\#1} shows how an envelope becomes active after the multi-party quorum is satisfied under mTLS-constrained identities; \textbf{Test\#2} shows deterministic \texttt{ALLOW/DENY} decisions using signed artifacts (ECT + DPoP) together with the referenced policy artifact;  \textbf{Test\#3} is for verify the governed MNIST prediction.  For each test, we summarize the required inputs, the observable outputs, and the expected failure modes so that the reader can validate the trust chain without following implementation details line by line.
 
\paragraph{Test\#1 — Envelope creation (quorum):} Run Test1A\_createEnvelope.sh, which:
\begin{enumerate}
    \item Initializes a bind with two participants and quorum \texttt{k=2}, \texttt{n=2}.
    \item Prompts the operator to enter the two \textit{6-digit} codes displayed by  two admin devices, posting /verify-start.
    \item Claims sessions via the hub credential (/session/claim) and submits approvals; quorum completion yields an envelope\_id.
\end{enumerate}
 
Successful completion prints:   Envelope created: <uuid> and suggests a post-envelope integration test hook (\texttt{./Test1B\_postEnvelope.sh \$ENVELOPE\_ID}). 
 
\paragraph{Test\#2 — Trust chain admission: ECT + DPoP (\texttt{/mint\_ect}, \texttt{/admission/check}).}
Run the trust-chain script harness (\texttt{Test2A\_run\_probe\_eddsa\_nginx.sh}), which:
\begin{enumerate}
    \item Generates a holder key pair (member key material).
    \item Mints an ECT via \texttt{/mint\_ect} under organization-admin mTLS, granting selected capability profiles (by reference to \texttt{cap\_profiles} in \texttt{policy.json}; in the script: \texttt{capset:trainer\_A} and \texttt{capset:data\_scientist}).
    \item Generates a DPoP proof (EdDSA) bound to the holder key and request parameters (method, HTU, nonce, JTI).
    \item Executes admission probes against \texttt{/admission/check} that must \texttt{ALLOW} and \texttt{DENY} deterministically. The \texttt{ALLOW} probe uses a clinical-style tuple vocabulary (e.g., \texttt{resource=TUMOR\_MEASUREMENTS}, \texttt{action=read}, \texttt{agg=aggregated}, \texttt{pii=false}, \texttt{contact=false}). The \texttt{DENY} probes cover (i) wrong purpose (e.g., \texttt{resource=PET-CT}, \texttt{action=train}, \texttt{purpose=model\_prediction}), (ii) wrong cohort (e.g., \texttt{cohort=B} when only \texttt{A} is permitted), and (iii) holder-binding mismatch by presenting the ECT with a DPoP generated under a different keypair (intruder).
\end{enumerate}
The test is considered successful only if the verifier behaves deterministically under repeated evaluation of the same inputs. Concretely, when presented with the same request tuple and the same signed artifacts, it must always return the same \texttt{ALLOW} or \texttt{DENY} outcome. It must also fail closed, i.e., if a required artifact is missing, if a signature is invalid, or if proof of possession is inconsistent with the holder binding encoded in the ECT, the result must be \texttt{DENY} without fallback or discretionary interpretation. This test is limited to the minting and verification path. The clinical attributes included in the request are not used to drive a real prediction workload; they are treated only as tuple fields against which the verifier matches the signed capability constraints. The purpose of the test is therefore not to measure model behavior, but to show that capability issuance and local verification produce stable, reproducible decisions from the same declared inputs.

\paragraph{Test\#3 — Governed MNIST prediction (clinical surrogate).}
The prediction test demonstrates the operational path \texttt{Hub $\rightarrow$ /admission/check $\rightarrow$ flower-server:/predict\_image}. MNIST digits serve as a stand-in for clinical imaging resources (e.g., PET-CT), and cohorts (\texttt{EVEN\_ONLY}, \texttt{ODD\_ONLY}, \texttt{ODD\_PLUS}) act as regulated “patient groups/study strata” to demonstrate scope enforcement. An issuer mints an ECT for a member with a declared cohort, and the user submits a prediction request (\texttt{Test3A\_run\_probe\_mnist.sh}). If admission is denied (e.g., due to a cohort mismatch or a capability violation), no prediction executes. If admitted, the prediction service additionally enforces cohort-scoped label disclosure by masking disallowed classes at inference time (\texttt{Test3B\_run\_predict\_via.sh}). An E2E CLI regression check (\texttt{Test3C\_e2e.sh}) completes the demonstration.

\paragraph{E2E UI demonstrator (admin mint $\rightarrow$ user governed predict).}
In addition to the CLI tests, the PoC provides a minimal web UI at \texttt{http://127.0.0.1:8082/} that follows the same governed execution path end-to-end. The UI follows the two-step protocol: (i) Select the \texttt{Admin Tab} to let an organization-scoped issuer (e.g., \texttt{issuer-hospitala} / \texttt{issuer-hospitalb}) mint an ECT for the selected member and cohort. In turn, it will call the verifier \texttt{/mint\_ect}; (ii) Select the \texttt{User tab} to let the user submit \{\texttt{who}, \texttt{envelope\_id}, \texttt{cohort}, \texttt{digit}\} along with the minted ECT (and DPoP) to the boundary. The hub then performs \texttt{/admission/check} and, only on \texttt{ALLOW=true}, forwards internally to \texttt{flower-server:/predict\_image}; denials return a structured reason (e.g., capability/cohort violation) and no execution is performed. The UI also makes operational readiness explicit if the referenced \texttt{envelope\_id} has no persisted model under \texttt{/vault/<envelope\_id>/model.pth}, prediction returns \texttt{model\_not\_ready}. This UI layer is intentionally thin. It demonstrates that FCaC’s boundary guarantees remain unchanged when driven by a human-facing interface, preserving separation of duties (issuers mint, hub admits/dispatches, backend stays internal-only).

\paragraph{Current PoC limitations}  
\begin{itemize}[topsep=0.5pt, leftmargin=1.8em]
    \item \textbf{L1. Envelope identifier not token-bound:} The \texttt{envelope\_id} is used by the service as a routing selector to load persisted envelope or model state, but it is not currently part of the minting or verification tuple.

    \item \textbf{L2. Set-scoped capabilities:} In the current implementation, capabilities are \emph{set-scoped}: an issued ECT grants the right to invoke a class of operations over any envelope that is \emph{Active} at verification time, where \emph{Active} is defined by an administrator-controlled validity window \texttt{[nbf, exp]}.

    \item \textbf{L3. Activity check rather than exact envelope match:} Verification enforces proof validity (ECT chain and DPoP binding) and checks that the requested \texttt{envelope\_id} is active at the current time, instead of requiring an exact identifier match embedded in the token.

\item \textbf{L4. Envelope-specific capabilities not yet evaluated:} Binding \texttt{envelope\_id} into the minting and verification tuple is straightforward and would yield envelope-specific capabilities, authorizing exactly one envelope or a bounded set. However, we have not yet evaluated issuance policies, lifecycle semantics, or benchmarks under this more granular model. This refinement would also simplify administrative reallocation: relocating a member would become a token-issuance update (minting a new envelope-scoped ECT and letting the old one expire or be revoked), rather than managing implicit access over the entire active set. This would reduce cross-envelope privilege and accidental operator drift.

    \item \textbf{L5. Cryptographic roles only partially separated in the PoC:} The conceptual split between organization root (KYO), issuer authority for ECT signing, and holder keys for proof of possession is clearer in the architecture than in the present prototype. The PoC exercises the trust chain end to end, but it does not yet fully model richer issuer and holder role separation as required in more mature multi-party deployments.

    \item \textbf{L6. Key management is simulated rather than hardware-backed:} The prototype simulates vault-backed storage for signing material on disk. Production deployments would use organization-controlled hardware-backed key management or enclave-backed signing services to protect issuer keys and holder credentials more strongly.

    \item \textbf{L7. Immediate revocation is not implemented:} The PoC supports bounded validity and key rotation, but not immediate revocation of a still-valid ECT or delegation link through out-of-band revocation evidence cached by verifiers.

    \item \textbf{L8. Scope of evidence is demonstrative rather than production-complete:} The PoC demonstrates deterministic verification, capability scoping, and end-to-end trust-chain behavior, but it does not yet evaluate larger-scale operational concerns such as concurrent multi-envelope issuance, richer delegation topologies, or production-grade performance under sustained multi-party load.
\end{itemize} 

 \section{Deployment Models}
\begin{figure}[ht]
    \centering
    \includegraphics[width=1.0\linewidth]{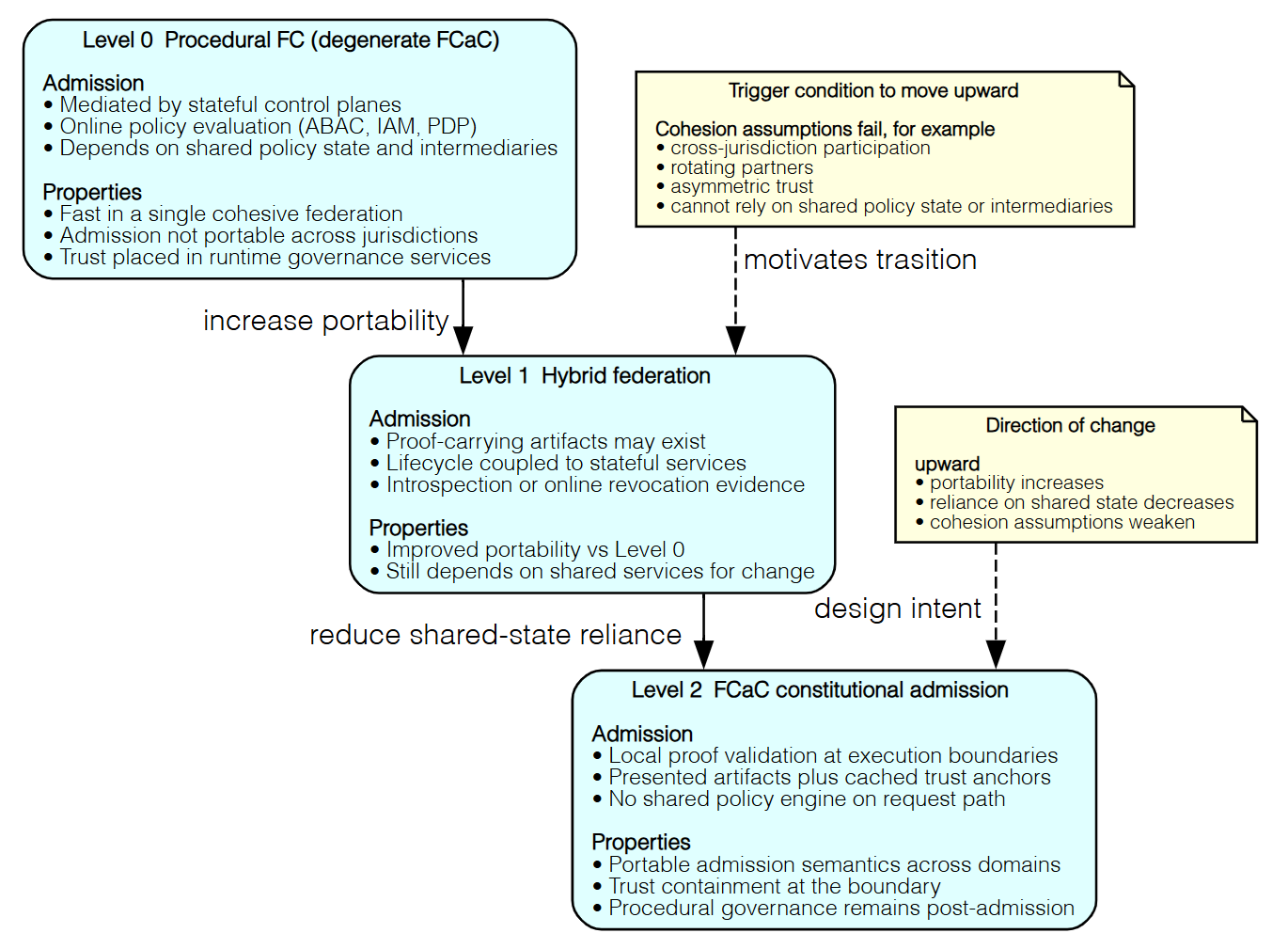}
    \caption{Deployment mode of federated governance regimes (FC $\to$ Hybrid $\to$ FCaC). Stateful-only federations form a limiting case (a degenerate FCaC) in which admission is mediated by stateful control planes. FCaC targets regimes where admission must remain portable across domains through locally verifiable artifacts.}
    \label{fig:governance}
\end{figure}

FCaC can be deployed in three principal regimes presented in Figure~\ref{fig:governance} that summarize under which conditions they are instantiated. In a single cohesive federation, shared control planes, aligned policy state, and institutional coordination may be sufficient to govern execution. FCaC is well suited when the cohesion assumptions fail. Typical trigger points include cross-jurisdiction participation, rotating partners, asymmetric trust, or the inability to rely on shared policy state and intermediaries. In such settings, local control alone does not yield portable admission semantics. 

\begin{itemize}[topsep=0.5pt, leftmargin=1.8em]
\item  \textbf{Level 0}: conventional FC deployments implement FC principles through shared control planes. Identity systems and policy evaluation decide whether requests may proceed, and enforcement depends on shared operational state and intermediaries. This regime supports rapid deployment in a single cohesive federation but does not provide portable semantics across jurisdictions. 
  
\item \textbf{Level 1}: hybrid federations introduce proof-carrying artifacts but still couple lifecycle decisions, such as revocation and delegation updates, to shared services. This improves portability but retains dependence on those services for lifecycle changes. 
    
\item \textbf{Level 2}: FCaC validates presented proof-carrying artifacts at execution boundaries using cached trust anchors. Control after boundary verification remains local and is administered independently within each domain.
\end{itemize}

FCaC complements existing identity systems, policy engines, and established control frameworks. In cohesive federations with a strong coordinating operator, FC principles can be realized through shared control planes, identity infrastructure, and local workflow enforcement. FCaC targets the stricter regime in which such shared policy state or intermediaries cannot be assumed. In that setting, cross-domain execution must be determined through portable, locally verifiable artifacts rather than through runtime dependence on a common control plane.

\section{Conclusion and Future Work}
This paper introduced Federated Computing as Code (FCaC), a declarative architectural discipline for enforcing sovereignty-critical constraints in federated systems through cryptographically verifiable artifacts. By structuring a federation as a Virtual Federated Platform (VFP) composed of Core, Business, and Governance contracts, FCaC makes permission to execute and delegate locally checkable at organizational boundaries, while preserving operational autonomy inside each participating domain. Boundary decisions are enforced through validation of verifiable artifacts rather than through online policy interpretation or centralized trust infrastructure.

A central contribution is the precise localization of enforcement. FCaC enforces sovereignty-critical constraints deterministically at execution boundaries through verification of proof-carrying artifacts. Controls that depend on context, local state, or workflow progression—such as ABAC, quotas, monitoring, and incident response—remain inside Core environments after the boundary check. This separation yields reproducible decisions at organizational and jurisdictional boundaries without requiring shared policy engines or institutional intermediaries.

FCaC does not attempt to mechanize obligations that depend on mutable global state, discretionary interpretation, or ex post judgment. It constrains execution only where requirements can be reduced to mechanically verifiable conditions enforced at the boundary; everything else remains with backend systems and organizational processes. This separation distinguishes FCaC from policy-centric and identity-centric approaches that attempt to subsume boundary decisions and operational control into a single enforcement plane.

Several directions for future work follow naturally. First, extending envelope creation to quorum-based or multi-authority regimes would support federations in which authority is jointly defined and not unilaterally issued. Second, richer delegation structures and lifecycle semantics should be explored to formalize and evaluate revocation mechanisms beyond the base lifecycle support already described (Section~\ref{ssec:krot}), and to explore temporal attenuation and bounded delegation. Third, a formal account of the compilation process from declarative contracts to verifiable artifacts would strengthen the architecture's theoretical grounding. Finally, broader empirical validation through additional proof-of-concept deployments, including large-scale federated learning and cross-sector data exchange, is needed to assess applicability and operational trade-offs.

FCaC delineates an architectural space in which sovereignty-critical boundary decisions become a verifiable property of execution. It provides a design discipline for federated systems that require portable, locally verifiable artifacts across independently governed domains.


\bibliographystyle{unsrturl}  
\bibliography{references}   
\newpage
\appendix
\section{RFCs used with FCaC}
\label{app}
\paragraph{RFC 8705 (OAuth 2.0 Mutual-TLS Client Authentication and Certificate-Bound Access Tokens)~\cite{rfc8705}:} Establishes organizational root of trust through X.509 certificates; binds access tokens to client TLS certificates, creating a verifiable chain back to a certificate authority;  anchors identity to organizationally verifiable credentials; enables \textit{Key Your Organization} (KYO) pattern for cross-jurisdictional trust.
 
\paragraph{RFC 7519 (JSON Web Token -- JWT)~\cite{rfc7519}:}
Defines a compact, URL-safe token format for representing claims as a signed or encrypted JSON object. A JWT consists of three parts: a header, a payload of claims, and a signature or integrity protection. In FCaC, JWT provides the basic container format used to carry capability claims. However, JWT does not guarantee proof of possession. If used as a bearer token, authorization depends only on presentation of the token instead of on cryptographic demonstration of key ownership.

\paragraph{RFC 7800 (Proof-of-Possession Key Semantics for JWTs)1\cite{rfc7800}:}
 Transforms bearer tokens into proof-of-possession tokens via the \textit{cnf} (confirmation) claim. Binds JWTs to cryptographic keys held by the presenter. Enables verification that the token presenter controls the associated private key. Critical for preventing token theft and replay attacks in capability-based systems.
 
\paragraph{RFC 7638 (JSON Web Key Thumbprint)~\cite{rfc7638}:}
 Provides a deterministic, canonical method for generating stable cryptographic key identifiers. Defines hash-based thumbprints of JWKs for compact key references in \textit{cnf.jkt} fields. Ensures interoperability through standardized computation. Essential for delegation chains and key references in distributed systems where embedding full keys is impractical.
 
\paragraph{RFC 9449 (OAuth 2.0 Demonstrating Proof of Possession - DPoP)~\cite{rfc9449}:}
 Binds token usage to specific HTTP requests at execution time. Requires fresh, request-specific JWT proofs signed with the bound private key. Includes HTTP method, URL, and timestamp to prevent replay attacks. Provides moment-of-execution verification that bridges abstract capabilities (what you can do) to concrete actions (what you're doing now).

\end{document}